\documentclass[aps,prb,floatfix,showkeys,amssymb,12pt]{revtex4}

\usepackage{graphicx}
\begin{document}

\baselineskip 0.8cm

\newcommand{\threeby}{3$\times$3}
\newcommand{\MnN}{Mn$_{3}$N$_{2}$}
\newcommand{\etaMnN}{Mn$_{3}$N$_{2}$}

\title{Scanning Tunneling Microscopy Study of Square Manganese Tetramers on \etaMnN (001)}
\author{Rong Yang, Haiqiang Yang, and Arthur R. Smith\footnote{Corresponding author, smitha2@ohio.edu}}

\affiliation{Nanoscale and Quantum Phenomena Institute,Department of Physics and Astronomy, Ohio University, Athens, OH
45701}

\begin{abstract}
\baselineskip 0.8cm
We have investigated the growth of antiferromagnetic \etaMnN(001) on MgO(001) by molecular beam epitaxy and
scanning tunneling microscopy . The images show smooth terraces and atomic steps. On some of the terraces a unique
and new reconstruction is seen, resolved as square Mn tetramers in a c(4$\times$2) structural arrangement. Two
domains of the tetramer reconstruction, rotated by 90$^\circ$ to each other, occur. A model is presented for this
square Mn tetramer reconstruction, in which the Mn atoms of the tetramer layer belong to the Mn layer at the
surface in the MnN-MnN-Mn stacking sequence.
\end{abstract}
\maketitle
\newpage
Reconstructions of many semiconductor surfaces have been explored intensively in the past several years due to
their importance in understanding the epitaxial growth process. \cite{StekolniKov,Feenstra,Kastner1,Smith1} The growth
of transition metal nitride layers has been a subject of significant interest recently due to their unique  structural,
electronic and magnetic properties.\cite{Suzuki1,Suzuki2} Due to the widespread interest inn ternary MnGaN,
\cite{Ohno1,Ohno2} it is crucial to also study the binary compounds. Moreover, due to their unique magnetic
properties, these compounds may themselves have useful technological applications.

We have previously investigated the epitaxial growth of \etaMnN\ on MgO(001).\cite{YangJAP,YangAPL}\etaMnN\ is a
layer-wise antiferromagnet with magnetic moments of $\sim$3$\mu_{B}$.\cite{Kreiner} Two orientations [(010) and
(001)] of this structure were grown controllably on MgO(001), depending on the growth conditions. The (010) surface
has been previously investigated in detail. Even the spin structure of this antiferromagnet was resolved using
Spin-polarized scanning tunneling microscopy (SP-STM).\cite{YangPRL,SmithSS} Although equally important so far very
little has been known about the (001) surface of \etaMnN\ .

In this letter, we present results for the (001) surface in which the Mn planes are parallel to the surface plane.
The orientation presents an interesting question regarding the epitaxial \etaMnN(001) growth: Will the surface
structure have a layer-dependence, similar to the bulk layer sequence (Mn-MnN-MnN-Mn$\ldots$)? This question is
addressed by studying the structure of the different terraces observed in the STM images. A new reconstruction is
found, which relates to the energy-lowering of the layer of Mn atoms. This reconstruction is explained using a
simple model.

The experiments are performed in a custom-designed ultra-high vacuum molecular beam epitaxy/scanning tunneling
microscopy(MBE/STM) system. The MBE system includes a solid source effusion cell for Mn and a RF plasma source for
N. To prepare the \etaMnN(001) surface, the MgO substrate is heated up to 1000 $^\circ$C for 30 minutes with the
nitrogen plasma on, and then the temperature is lowered to 450 $^\circ$C prior to the growth of manganese nitride.
The nitrogen flow rate is about 1.1 sccm (growth chamber pressure is 1.1$\times$10$^{-5}$ Torr) with the RF power
set at 500 W. The Mn flux is about 1.4$\times$10$^{14}$/cm$^{2}$s. The growth condition is determined from a
previous study to result in the epitaxial growth of \etaMnN(001). \cite{YangJAP} The surface structure is monitored
using reflection high energy electron diffraction (RHEED) during growth. Following growth, the samples are
transferred directly to the {\it in-situ} STM and STM measurements are performed at 300K.

Fig.1(a) shows a large image of size of 800 \AA $\times$ 800 \AA . This growth surface indeed shows large smooth
terraces with steps. Different terraces show different structural features. Some terraces look uniform, while
others have some defects. Since the layer sequence is Mn-MnN-MnN$\ldots$,\cite{YangAPL,Kreiner} it is expected that
some terraces correspond to Mn layers, while others correspond to MnN layers. We expect to see 2 layers having same
structure, and 1 layer having different structure. The layer with some defects is labeled as A. The layer under it
as B1, the layer above it as B2. The terraces in the image are labeled in this way. By measuring the height
difference between different terraces, we deduced that the layer with some defects is Mn layer-referred to as layer
A. The layer under it  is therefore MnN layer, the layer above it is also MnN layer.

Shown in Fig.1(b) are single line-profiles of line 1 and line 2 labeled in Fig.1(a). With this layer assignment, by
analyzing the step heights, the step height between layer A and B1 is h1=$\mid$A$\rightarrow$B1$\mid$=1.33$\pm$0.01
\AA . The step height between layer B1 and B2 is h2=$\mid$B1$\rightarrow$B2$\mid$=2.15$\pm$0.01 \AA, and the step
height between layer B2 and A is h3=$\mid$B2$\rightarrow$A$\mid$=2.41 $\pm$0.01 \AA . The total z-height of one
period A-B1-B2-A is 5.90 $\pm$0.03\AA, which is  very close to, but less than, the bulk value of lattice constant
c/2 ( c/2=6.07 \AA )\cite{YangJAP} of \etaMnN ( z calibration has been performed on other samples). This can be
attributed to a surface relaxation, in which the outer surface layers of the film relax inwards by, in this case,
-2.8$\%$.

Shown in Fig.1(c) is a schematic surface side-view model of \etaMnN(001) grown on MgO(001). It is fct and has
octohedral bonding. Along the growth direction, the structure is the repeating sequence Mn-MnN-MnN-Mn-MnN-MnN
$\ldots$ (or A-B1-B2-A-B1-B2$\ldots$).The image and the model agree well with each other since both
show the different structures on different terraces.

Fig.2 shows STM images of the Mn (A) layer of \etaMnN(001).  In Fig 2(a), nanometer-size protrusions are observed
which are arranged in a pattern similar to a hexagon; but this is not expected for a cubic (001) surface, so at
first it appears difficult to explain. But, in Fig.2(b), higher resolution shows that each protrusion in Fig.2(a)
can be resolved into 4 smaller dots of atom size. Now, it is seen that the 4-atom units are aligned along straight
orthogonal rows. Since these rows can belong to one of 2 possible domain types (marked $\it{1}$ and $\it{2}$),
which lie at 90$^\circ$ to each other, we see that the cubic symmetry of the surface is manifested.

Since the spacing of the 4-atom unit rows is 2$a$ ($a$=$a_{0}/\sqrt{2}$=2.98 \AA, which is the spacing of two
adjacent Mn atoms on the 1$\times$1 MnN surface ), the rows could shift by half their spacing. This is observed at
the point marked $\it{S}$ in the image [Fig.2(b)](compare the 2 sets of drawn lines). We also notice, in addition
to the domains, several areas which appear to be defects ( marked by region $\it{3}$ ).

Next, we measured the spacings between the 4-atom units along different directions. The spacing between two units
along [$\bar{1}$10] is $\sim$ 11.9 \AA , which is 4 times the spacing of two adjacent Mn atoms on the 1$\times$1
MnN terrace surface along [$\bar{1}$10]. The spacing between two units along [110] axis is $\sim$ 5.97 \AA\ , which
equals 2$a$. Clearly, this ordering forms a surface symmetry with c(4$\times$2) reconstruction periodicity.

The whole structure can be explained as a regrouping of a full monolayer of Mn atoms into 4-atom units - Mn
"tetramers". The tetramers naturally form into rows, but due to the manner of the re-grouping, which alternates
between adjacent rows, the tetramers do not take on a simple square lattice, but rather a distorted hexagonal
lattice of tetramer units appears. Because of this the unit cell is not 2$\times$2, but rather c(4$\times$2).
Moreover, the 4-fold symmetry naturally results in 2 types of rotated domains. The defect structures are now
understood as regions of missing Mn atoms or Mn tetramers.

Fig.3 shows zoom-in STM images of the 1$\times$1 MnN (B1) layer [Fig.3(a)] and the c(4$\times$2) Mn(A) layer
[Fig.3(b),3(c)]. Fig.3(a) shows an atom-resolved MnN surface with 1$\times$1 periodicity along [$\bar{1}$10] and
[110] directions. The imaged atoms are the Mn atoms;  From recent theoretical caculations, N atoms have much lower
density of states(DOS) compared with Mn in MnN \cite{Walter}. Fig.3(b) shows an image of the c(4$\times$2) tetramer reconstruction of the Mn atom
layer. Fig.3(c) shows also the c($4\times$2)reconstruction, but with higher resolution, revealing clearly the 4
atoms composing each Mn tetramer. The corresponding structural models for the images are shown in Fig.3(d) and
Fig.3(e) for the 1$\times$1 structure of the MnN layer and for the c(4$\times$2) reconstruction of the Mn layer,
respectively.

To explain the formation of Mn tetramers, a side-view, bulk-terminated model is shown in Fig.4. Since the Mn
surface atoms in the top layer are bonded atop the N atoms of the B1-MnN layer, it is expected that this Mn
monolayer will not be stable in this configuration.  To achieve a more stable and energetically- favorable
structure, it is reasonable that 4 neighboring Mn atoms be slightly displaced towards each other, to form
tetramers, as shown in Fig.4. The 4 Mn atoms may thus form intra-tetramer bonds, the attractive nature of which may
be balanced by the resultant stress from the tilted N-Mn bonds. This agrees with the experiment because under certain
tip conditions, the Mn tetramers appear as a single protrusion. We also noticed that the step height measurements
show the A-B1 step height as ~1.33 \AA\ at $V_{s}=0.6V$, smaller than the bulk interlayer spacing of 2.02 \AA. This
is completely consistent with a slight inward relaxation of the Mn atoms and/or a redistribution of the DOS
spatially due to the intra-tetramer bond formation. We conclude that the bulk-terminated Mn atom layer is
structurally unstable, and formation of Mn tetramers occurs to stabilize the layer.

In conclusion, we have used STM to study the surface reconstructions of \etaMnN(001) surface on MgO(001). Two
different structures are observed in different terraces. There is a 1$\times$1 structure of the MnN layer, and a new
reconstruction is seen in the Mn layer, resolved as square Mn tetramers with a conventional unit cell c(4$\times$2).
Two types of reconstruction domains rotated by 90$^{0}$ to each other are also observed. A
schematic surface model has been presented which has the correct translational periodicity and rotational symmetry,
and which agrees well with the observed features of the STM images. A model has also been proposed for explaining this
square Mn tetramer reconstruction within the Mn layer at the surface. The tetramer formation is attributed to the
energy lowering due to the re-bonding of the Mn layer.

\newpage

\begin{figure}
\includegraphics[width=7cm]{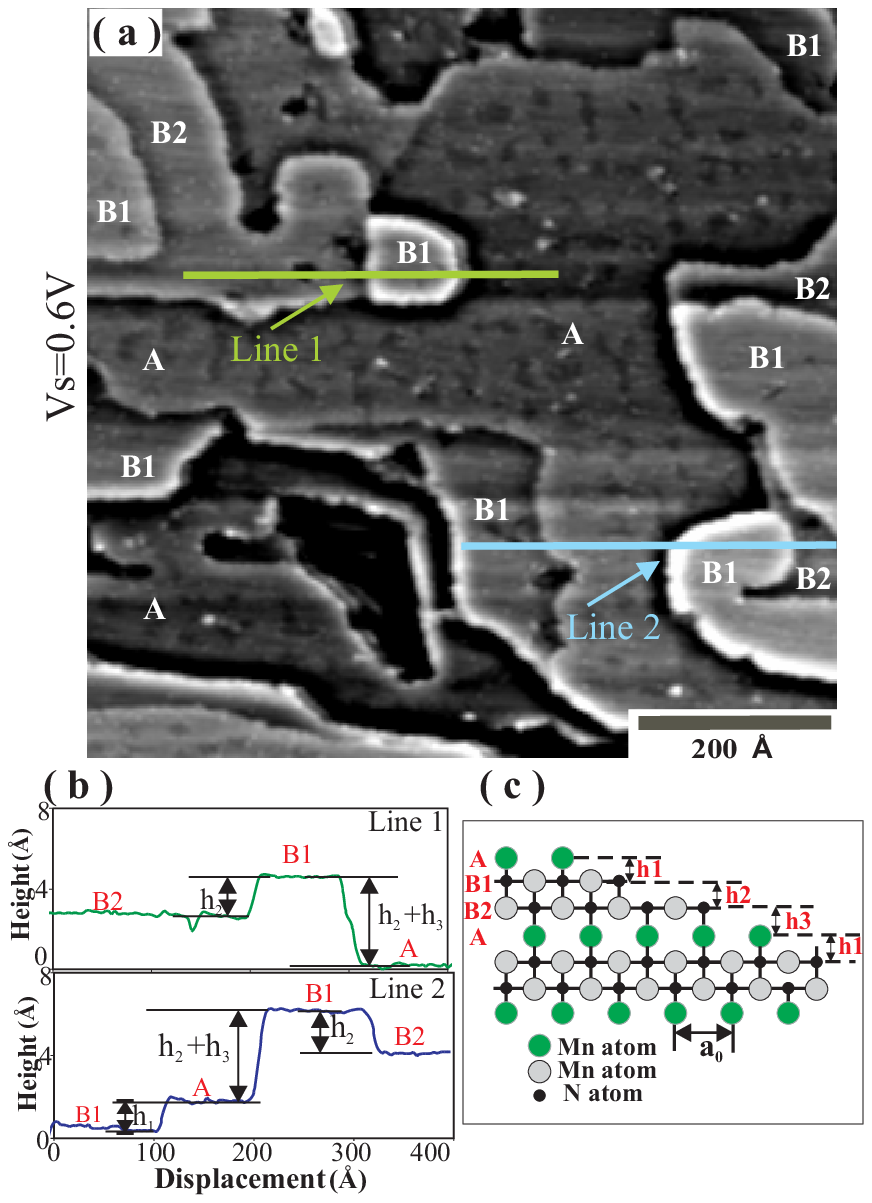}
\caption{STM images of \etaMnN(001), single line-profiles and a surface side-view model are shown. (a) A STM image
of 800 \AA $\times$ 800 \AA\ ( V$_{s}$=0.6 V and I$_{t}$=0.8 nA ). The surface stacking sequence Mn-MnN-MnN is
labeled as A, B1 and B2. (b) Height profiles of line 1 and line 2. The step height
h1=$\mid$A$\rightarrow$B1$\mid$=1.33$\pm$0.01 \AA , h2=$\mid$B1$\rightarrow$B2$\mid$=2.15$\pm$0.01 \AA,
h3=$\mid$B2$\rightarrow$A$\mid$=2.41$\pm$0.01 \AA. (c) Surface side-view model of \etaMnN(001). $a_{0}$=4.21\AA.}
\end{figure}

\newpage
\begin{figure}
\includegraphics[width=7cm]{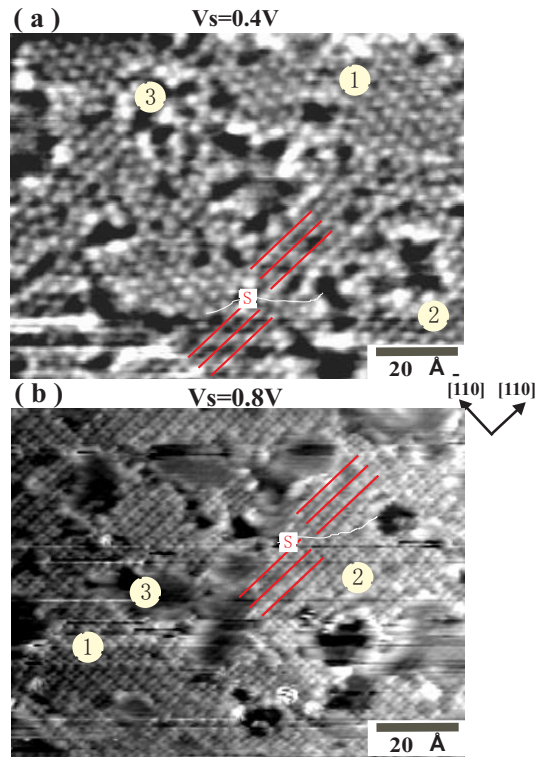}
\caption{STM images of the \etaMnN(001) c(4$\times$2) reconstruction. (a)Apparent "hexagonal" surface structure of
c(4$\times$2)surface(V$_{s}$=0.4V and I$_{t}$=1.0 nA, gray scale range is 0.5 \AA ). (b) Resolution of the detailed
square tetramers with orthogonal row structure of c(4$\times$2) reconstruction(V$_{s}$=0.8V and I$_{t}$=0.8nA,
 gray scale range is 1.28 \AA ).}
\end{figure}

\newpage
\begin{figure}
\includegraphics[width=7cm]{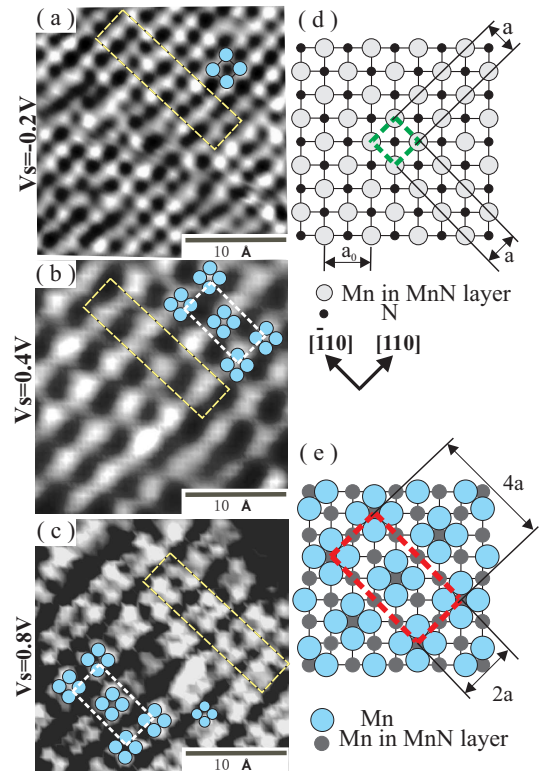}
\caption{Zoom-in STM images of the 1$\times$1 and c(4$\times$2) reconstructions, height profiles and corresponding
surface models. V$_{s}$ for three images in (a),(b),(c) are -0.2 V, 0.4 V and 0.8 V; and I$_{t}$ are 1.0 nA, 1.0 nA
and 0.8 nA, respectively ; And gray scale ranges are 0.5\AA, 0.5\AA\ and 1.1 \AA, respectively. (d) Model of
1$\times$1 MnN surface structure; (e) Model of c(4$\times$2) Mn atom reconstruction.}
\end{figure}

\newpage
\begin{figure}
\includegraphics[width=7cm]{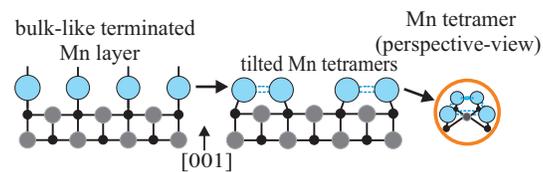}
\caption{Side-view schematic models of the Mn tetramers.}
\end{figure}

\end{document}